\documentclass[floats,prd,aps,amssymb,nofootinbib,showkeys]{revtex4}
\usepackage{amssymb}
\usepackage{graphics}
\usepackage{amsmath}
\usepackage{xcolor}
\usepackage{amsfonts}
\usepackage{bm}
\usepackage[]{latexsym}

\newcommand{\be}{\begin{equation}}\newcommand{\ee}{\end{equation}}
\newcommand{\bea}{\begin{eqnarray}}\newcommand{\eea}{\end{eqnarray}}
\newcommand{\brr}{\begin{array}}\newcommand{\err}{\end{array}}
\newcommand{\bit}{\begin{itemize}}\newcommand{\eit}{\end{itemize}}
\newcommand{\ben}{\begin{enumerate}}\newcommand{\een}{\end{enumerate}}

\newcommand{\ba}{\begin{array}}
\newcommand{\ea}{\end{array}}

\def\lf{\left}

\def\1{{_{1}}}\def\2{{_{2}}}

\def\noHe0{:\;\!\!\;\!\!:H_e(0):\;\!\!\;\!\!:}
\def\noHm0{:\;\!\!\;\!\!:H_\mu(0):\;\!\!\;\!\!:}

\newcommand{\dr}{\mathrm{d}}

\def\lf{\left}

\def\1{{_{1}}}\def\2{{_{2}}}

\def\lf{\left}

\def\1{{_{1}}}\def\2{{_{2}}}

\begin{document}
\title{Gravitational effects on neutrino decoherence in Lense-Thirring metric}

\author{Giuseppe Gaetano Luciano\footnote{email: gluciano@sa.infn.it (corresponding author)}$^{\hspace{0.1mm}1,2}$ and Massimo Blasone\footnote{email: blasone@sa.infn.it}$^{\hspace{0.1mm}1,2}$} 
\affiliation
{$^1$Dipartimento di Fisica, Universit\`a di Salerno, Via Giovanni Paolo II, 132 I-84084 Fisciano (SA), Italy.\\ 
$^2$INFN, Sezione di Napoli, Gruppo collegato di Salerno, Via Giovanni Paolo II, 132 I-84084 Fisciano (SA), Italy.}

\date{\today}

\begin{abstract}
We analyze gravity effects on neutrino wave packet decoherence. 
As a specific example, we consider the gravitational field of a spinning spherical body described by the Lense-Thirring metric. By working in the weak-field limit and employing Gaussian wave packets, we show that the characteristic coherence length of neutrino oscillation processes is nontrivially affected, the corrections being dependent on the mass and angular velocity of the gravity source. Possible experimental implications are finally discussed. 
\end{abstract}


\date{\today}

\maketitle

\section{Introduction}
Neutrinos are among the elementary particles
in the Standard Model (SM) of fundamental interactions. 
In spite of this, their essential nature has not yet been fully revealed, 
becoming even more puzzling after Pontecorvo's pioneering idea  
of neutrino mass and mixing \cite{Pontec1,Pontec4,Pontec2} and the subsequent 
discovery of flavor oscillations \cite{OscExp1,OscExp2,OscExp3,OscExp4}.
Further studies in Quantum Field Theory (QFT) have highlighted the shortcomings 
of the standard quantum mechanical (QM) predictions 
by pointing out the unitary inequivalence
between the Fock spaces for definite flavor fields and definite mass fields \cite{BV95,Caplup,Ji}. Phenomenological implications of this inequivalence have 
been investigated in a variety of contexts, ranging from vacuum effects
\cite{Casimir,Unruh,Unruhbis} to particle decays \cite{Protdec1,Protdec2,Protdec3,Protdec4} and apparent violations of the
weak equivalence principle~\cite{WeP}. 

Neutrino mixing and oscillations are typically analyzed in the plane wave approximation.  
However, a more realistic treatment that accounts for 
neutrinos being localized particles should involve the use
of wave packets (WP), which introduce decoherence among  
the mass eigenstates. The first WP approach was developed in \cite{Nussinov}, showing the existence of a coherence length beyond which the interference between massive neutrinos becomes negligible. 
This effect arises from the different group velocities of the different mass states, which 
leads WPs to spread over macroscopic sizes and separate during the propagation. Wave packet models of neutrino oscillations were later developed within the framework of QM \cite{Giunti1,Giunti2,Dolgov} and QFT \cite{QFTWP1,QFTWP2,QFTWP3,QFTWP4}, both in vacuum and matter \cite{Matter1, Matter2, Matter3} (see \cite{GiuntiWP} for a review). In particular, in dense environments decoherence by WP separation was shown to depend on the model chosen for the adiabaticity violation
of WP evolution \cite{Matter2,Matter3}.

All of the above investigations have been performed
in flat spacetime. Gravity effects on neutrino decoherence
were addressed in \cite{Volpe} considering a static and spherically symmetric field described by the Schwarzschild metric. 
By adopting the density matrix formalism \cite{Matter3} with Gaussian WPs and exploiting previous achievements of \cite{Stodo,Cardall,Fornengo},  neutrino decoherence was quantified by a coherence coordinate distance and a proper time. As a result, it was shown that these quantities are nontrivially
modified with respect to the flat case, the corrections being in principle
sizable. 

The Schwarzschild solution provides a first useful
approximation to describe the spacetime metric
around many astronomical objects, including the
Earth and Sun. However, it does not account for the rotation
of the source. In this work we take a further step forward by 
studying neutrino decoherence in the gravitational field
of a spinning spherical body of constant density and in the
weak-field regime (Lense-Thirring metric) \cite{Lense}. 
Following \cite{Volpe}, we resort to the density matrix approach and
evaluate the coherence length in terms
of the neutrino local energy, which is the energy actually
measured by a inertial observer at rest at finite radius in the gravitational field. 
In this sense, our calculation differs from that of \cite{Volpe}, 
where the final result is exhibited as a function of the 
asymptotic energy of neutrinos. 
We show that it is possible to extract a separate
gravitational contribution depending on the mass
and angular velocity of the source. Experimental 
implications of our finding  are preliminarily discussed at the end. 

The layout of the paper is as follows: in Sec. \ref{Sec1} we
review the density matrix approach to describe neutrino WP decoherence in flat spacetime. For this purpose, we follow \cite{Matter3,Volpe}. The above considerations are then extended to curved spacetime in Sec. \ref{Sec2}. As a specific example, we consider neutrino propagation in 
the Lense-Thirring metric. Conclusions and outlook are summarized in Sec. \ref{Sec3}. Throughout all the manuscript, we use natural units $\hbar=c=G=1$ and the metric with the conventional mostly negative signature $(+1,-1,-1,-1)$.

\section{WP decoherence in flat spacetime: the density matrix approach}
\label{Sec1}
In the SM it is a well-established fact that neutrinos
interact in weak eigenstates that are superpositions
of mass eigenstates through Pontecorvo transformation \cite{Pontec1,Pontec4,Pontec2}
\begin{equation}
\label{flavor}
|\nu_\ell(t,\vec{p})\rangle=\sum_{i=1,2}U^*_{\ell\hspace{0.1mm}i}\hspace{0.2mm}|\nu_i(t,\vec{p})\rangle\,,
\end{equation}
where $\ell=e, \mu$ ($i=1,2$) denotes the flavor (mass)
index and $U^*_{\ell \hspace{0.1mm} i}$ is the generic element
of Pontecorvo mixing matrix\footnote{We consider a simplified model
involving only two generations of neutrino. The same considerations
and results hold in the case of three flavors.}. 
The time-dependent neutrino state $|\nu_i(t,\vec{p})\rangle$ with three-momentum $\vec{p}$ is solution of the Schr\"odinger-like equation
$i \frac{d}{dt}|\nu_i(t,\vec{p})\rangle=H|\nu_i(t,\vec{p})\rangle$,
where $H$ is the Hamiltonian governing
the time evolution. In astrophysical environments the hamiltonian may include
different contributions, such as the vacuum term, the matter-
and self-interactions. However, following \cite{Volpe}, in our analysis 
we neglect matter effects and neutrino self-interactions outside the compact object. We are then left with $H_0|\nu_i(t,\vec{p})\rangle=E_i(\vec{p})|\nu_i(t,\vec{p})\rangle$, where $E_i(\vec{p})=(m_i^2+|\vec{p}|^2)^{1/2}$ is the free energy eigenvalue of the $i$-th mass eigenstate. 

In the coordinate space the $i$-th neutrino state is given by the
$3$-dimensional Fourier expansion 
\be
\label{coordsp}
|\nu_i(t,\vec{x})\rangle=\frac{1}{(2\pi)^3}\int d^3p\, e^{i\vec{p}\cdot\vec{x}}|\nu_i(t,\vec{p})\rangle\,.
\ee
To streamline the notation, henceforth we denote the momentum integration
$(2\pi)^{-3}\int d^3p$ by $\int_{\vec{p}}$\,.

In the standard treatment of flavor oscillations, 
the mass eigenstates are typically described by plane waves.
To account for the localization of neutrinos to a finite region,  
a formalism based on WPs should be used. This has
been considered in \cite{Nussinov,Giunti1,Giunti2,Dolgov,QFTWP1,QFTWP2,QFTWP3,QFTWP4,Matter1,Matter2,Matter3}. 
In the WP approach, the neutrino flavor state \eqref{flavor}
turns out to be a superposition of mass eigenstate WPs, 
each centered around the momentum $\vec{p}_i$ with 
distribution amplitude $f_{\vec{p}_i}(\vec{p})$. 
At the initial time $t_0=0$, the $i$-th WP component satisfies 
\be
|\nu_i(0,\vec{p})\rangle=f_{\vec{p}_i}(\vec{p})|\nu_i\rangle,
\ee
where $\langle\nu_j|\nu_i\rangle=\delta_{ij}$
and 
\be
\int_{\vec{p}} |f_{\vec{p}_i}(\vec{p})|^2=1\,.
\ee
The explicit form of the WP distribution $f_{\vec{p}_i}(\vec{p})$
will be given below. 

Now, by use of Eq. \eqref{coordsp},
the $i$-th mass state in the coordinate space
can be written as
\be
|\nu_i(t,\vec{x})\rangle=\psi_i(t,\vec{x})|\nu_i\rangle\,,
\ee
where the wave function $\psi_i(t,\vec{x})$ in the coordinate space
is the Fourier transform of the corresponding momentum-dependent wave function, i.e.
\be
\label{FT}
\psi_i(t,\vec{x})=\int_{\vec{p}}e^{i\vec{p}\cdot\vec{x}}\psi_i(t,\vec{p})
=\int_{\vec{p}}e^{i\vec{p}\cdot\vec{x}} f_{\vec{p}_i}(\vec{p})e^{-iE_i(\vec{p})t}\,.
\ee
Therefore, the flavor state takes the form
\be
\label{flafinst}
|\nu_\ell(t,\vec{x})\rangle=\sum_{i=1,2}U^*_{\ell\hspace{0.1mm}i}\hspace{0.2mm}\psi_i(t,\vec{x})|\nu_i\rangle\,. 
\ee

Starting from the above premises, let us employ the density matrix formalism
to describe WP decoherence effects. In flat spacetime, 
the one-body density matrix for the neutrino state \eqref{flafinst}
is defined as usual by \cite{Matter3}
\be
\label{densmat}
\rho^{(\ell)} (t,\vec{x})=|\nu_\ell(t,\vec{x})\rangle\langle \nu_\ell(t,\vec{x})|\,. 
\ee
By using Eqs. \eqref{FT} and \eqref{flafinst}, 
the $jk$-element of this matrix can be easily computed, yielding
\be
\label{jkelem}
\rho^{(\ell)}_{jk}(t,\vec{x})=U^*_{\ell\hspace{0.1mm}j} U_{\ell\hspace{0.1mm}k}\hspace{0.2mm}\psi_j(t,\vec{x})\psi_k^*(t,\vec{x})\,.
\ee

Flavor vacuum oscillations occur due to the interference
between different massive neutrinos. Thus, in the WP approach the condition to be satisfied to detect oscillations at a given point is that the mass eigenstate WPs still overlap sufficiently to produce interference in that point (see Fig. \ref{fig1}). To take account of this, one introduces the \emph{coherence length} $L_{\mathrm{coh}}$, which is defined 
as the distance travelled by neutrinos beyond which the WPs corresponding to
different propagation eigenstates composing the produced
neutrino flavor state separate by more than the WP size $\sigma_x$. 
Recalling that the decoherence
is generated by the different group velocities of the different 
mass eigenstate WPs, the coherence length can be estimated heuristically as 
$L_{\mathrm{coh}}\simeq \sigma_x v_g (\Delta v_g)^{-1}$ \cite{Matter3}, where $v_g$ 
is the average group velocity
of the WPs and $\Delta v_g$ is the difference between
the group velocities of the mass eigenstate WPs. 
For relativistic neutrinos 
we have \cite{Matter3, Volpe}
\be
\label{cohle}
L_{\mathrm{coh}}\simeq \frac{2E^2}{|\Delta m^2_{jk}|}\sigma_x\,,
\ee
where $\Delta m^2_{jk}=m^2_j-m_k^2$ and $E\simeq |\vec{p}|$
is the average energy between the interfering mass eigenstates. 
Clearly, in the plane wave approximation we have $L_{\mathrm{coh}}\rightarrow\infty$ 
because of the infinite spatial extension of plane waves. 

The above relation is to be compared with the
characteristic oscillation length in the plane wave
formalism, which is
\be
L_{\mathrm{osc}}\simeq\frac{4\pi E}{|\Delta m_{jk}^2|}\,.
\ee
Thus, it is evident the central r\^ole of the finite WP-width assumption 
in producing decoherence between the different 
neutrino eigenstate WPs. 

It is worth noting that in the presence of three neutrino generations, one should define a coherence length for each pair
of propagation eigenstates. In that case, complete decoherence occurs when the distance travelled by neutrinos is higher than all the coherence lengths. As remarked in \cite{Matter3}, partial decoherence may also be of interest in some physical contexts. 

We now aim at deriving the coherence length \eqref{cohle}  more rigorously  
by using the density matrix approach. As we shall see in the next Section, 
this formalism is also well-suited to be extended to curved spacetime.
Toward this end, we need to specify the explicit form of WPs. 
There exist in the literature several examples of
WPs, such as square or sech WPs. 
Here, we resort to the most common Gaussian WPs
of momentum width 
\be
\sigma_p=\frac{1}{2\sigma_x}\,,
\ee
for which
the (normalized) distribution amplitude $f_{\vec{p}_i}(\vec{p})$
reads
\be
\label{fWP}
f_{\vec{p}_i}(\vec{p})={\left(\frac{2\pi}{\sigma_p^2}\right)}^{3/4} e^{-\frac{\left(\vec{p}-\vec{p}_i\right)^2}{4\sigma_p^2}}\,.
\ee
Notice that the plane wave limit is recovered for $\sigma_p\rightarrow0$ (i.e. for $\sigma_x\rightarrow\infty$), which 
yields $f_{\vec{p}_i}(\vec{p})=[(2\pi)^3/\sqrt{V}]\,\delta^3\left(\vec{p}-\vec{p}_i\right)$, 
where $V$ is the normalization volume. 

Now, by plugging Eq. \eqref{fWP} into \eqref{jkelem}, we are led to
\be
\label{integ}
\rho^{(\ell)}_{jk}(t,\vec{x})=N^{(\ell)}_{jk}\int_{\vec{p},\vec{q}}e^{-i\phi_{jk}(t,\vec{x})}\, e^{-\left[\frac{(\vec{p}-\vec{p}_j)^2}{4\sigma_p^2}\,+\,\frac{(\vec{q}-\vec{q}_k)^2}{4\sigma_p^2}\right]}\,,
\ee
where 
\be
\label{phaseneutrsh}
\phi_{jk}(t,\vec{x})=[E_j(\vec{p})-E_k(\vec{q})]t\,-\,(\vec{p}-\vec{q})\cdot \vec{x}
\ee 
is the standard QM phase shift and
\be
\label{norm}
N^{(\ell)}_{jk}\equiv{\left(\frac{2\pi}{\sigma_p^2}\right)}^{3/2}U^*_{\ell j}\hspace{0.2mm}U_{\ell k}\,.
\ee
We have assumed equal dispersion
$\sigma_p$ for the mass eigenstate WPs. 

The integrals in Eq. \eqref{integ} can be computed by expanding the neutrino energy $E_j(\vec{p})$ around the WP central momentum $\vec{p}_j$ according to \cite{Volpe}
\be
\label{Eexpand}
E_j(\vec{p})\simeq E_j+(\vec{p}-\vec{p}_j)\cdot \vec{v}_j\,, 
\ee
(and similarly for $E_k(\vec{q})$), where $E_j\equiv E_j(\vec{p}_j)$
and $\vec{v}_j=\frac{\partial E_j}{\partial p}|_{\vec{p}=\vec{p}_j}$ is the group velocity of the $j$-th mass eigenstate. 
The relation \eqref{integ} then becomes
\begin{equation}
\rho^{(\ell)}_{jk}(t,\vec{x})=N^{(\ell)}_{jk}\int_{\vec{p},\vec{q}}
e^{-i(E_j-E_k)t}\,
e^{-i[(\vec{p}-\vec{p}_j)\cdot \vec{v}_j-(\vec{q}-\vec{q}_k)\cdot \vec{v}_k]t}\,
e^{i(\vec{p}-\vec{q})\cdot\vec{x}}\,e^{-\left[\frac{(\vec{p}-\vec{p}_j)^2}{4\sigma_p^2}\,+\,\frac{(\vec{q}-\vec{q}_k)^2}{4\sigma_p^2}\right]}\,.
\end{equation}
By explicit calculation of the Gaussian integrals, we get
\be
\label{Gaintsol}
\rho^{(\ell)}_{jk}(t,\vec{x})=\frac{N^{(\ell)}_{jk}}{(2\sqrt{\pi}\sigma_x)^6}
e^{-i(E_{jk}t-\vec{p}_{jk}\cdot \vec{x})}\,
e^{-\left[\frac{(\vec{x}-\vec{v}_{j}t)^2}{4\sigma_x^2}+\frac{(\vec{x}-\vec{v}_{k}t)^2}{4\sigma_x^2}\right]}\,,
\ee
where 
\be
\label{diffen}
E_{jk}\equiv E_j-E_k\,,\quad \, \vec{p}_{jk}\equiv \vec{p}_j-\vec{q}_k.
\ee

Let us now evaluate the
density matrix averaged over time
\be
\label{Tint}
\rho^{(\ell)}_{jk}(\vec{x})\equiv\int dt\, \rho^{(\ell)}_{jk}(t,\vec{x})\,, 
\ee
which 
is the typical quantity of interest in oscillation experiments. 
Tedious but straightforward calculations lead to 
\be
\label{factorflat}
\rho^{(\ell)}_{jk}(\vec{x})=A^{(\ell)}_{jk}\,\rho_{jk}^{\mathrm{osc}}(\vec{x})\,\rho_{jk}^{\mathrm{damp}}(\vec{x})\,,
\ee
where
\begin{eqnarray}
\label{AMP}
A^{(\ell)}_{jk}&\hspace{-1mm}=\hspace{-1mm}&\frac{U^*_{\ell j}U_{\ell k}}{\sqrt{2}\pi \hspace{0.4mm}v\hspace{0.4mm}\sigma_x^2}\,e^{-\frac{(E_{jk}\sigma_x)^2}{v^2}}\,,\\[2mm]
\label{Osc}
\rho_{jk}^{\mathrm{osc}}(\vec{x})&\hspace{-1mm}=\hspace{-1mm}&e^{i\left(\vec{p}_{jk}-\frac{2E_{jk}\vec{v}_g}{v^2}\right)\cdot\vec{x}}\,,\\[2mm]
\rho_{jk}^{\mathrm{damp}}(\vec{x})&\hspace{-1mm}=\hspace{-1mm}&e^{-\frac{(\vec{v}_j-\vec{v}_k)^2}{4v^2\sigma_x^2}x^2}\,. 
\label{damp}
\end{eqnarray}
The first term does not affect oscillations. The second one
is the oscillation term with the extra factor $2E_{jk}\vec{v}_g/v^2$, 
where we have set $\vec{v}_g\equiv(\vec{v}_j+\vec{v}_k)/2$ and $v\equiv(v_j^2+v_k^2)^{1/2}$. 
The last factor is the damping term, which is actually responsible for decoherence. 
In passing, we remark that the above averaged density
matrix can be alternatively calculated as a function of time by integrating Eq. \eqref{Gaintsol} over space coordinates. This has been done in \cite{Matter3}, obtaining a similar expression for the decoherence term through the identification $t\simeq|\vec{x}|$. 

From Eq. \eqref{damp}, we can now estimate the coherence length 
between the mass eigenstate WPs as the distance at which
the density matrix is suppressed by a factor $e^{-1}$. This gives \cite{Volpe}
\be
\label{cohleDM}
L_\mathrm{coh}=\frac{2\hspace{0.4mm}v\hspace{0.4mm}\sigma_x}{|\vec{v}_j-\vec{v}_k|}\simeq\frac{4\sqrt{2}E^2}{|\Delta m^2_{jk}|}\sigma_x\,, 
\ee
which agrees with the heuristic estimate in Eq. \eqref{cohle}, up to a numerical factor.

In the next Section we show how the density matrix formalism
is modified when extended to curved spacetime. In particular, 
we consider the case of Lense-Thirring metric and evaluate
gravity-induced corrections to the coherence length \eqref{cohleDM}.

\section{Gravity effects on WP decoherence: the Lense-Thirring metric example}
\label{Sec2}
Gravitational effects on neutrino oscillations have
been extensively analyzed in the literature
by using several approaches, e.g. 
the plane wave method \cite{Fornengo,Konno} or geometric 
formalisms \cite{Cardall,Zhang}
and in different metrics, such as Schwarzschild \cite{Fornengo,Burgard}, Kerr \cite{Wudka,Konno}, 
Friedmann-Robertson-Walker \cite{Visinelli} and Lense-Thirring \cite{Lambiase} metrics, among others. Recently, nontrivial
results have been obtained in extended theories
of gravity \cite{ETG}, quantum gravity scenarios \cite{Mavromatos,Nicolini} 
and in QFT on Rindler
background \cite{Unruh}. 

The first systematic WP treatment of decoherence in 
neutrino oscillations has been developed in \cite{Volpe} 
by relying on the Stodolsky's covariant generalization of the quantum 
mechanical phase shift in Eq. \eqref{phaseneutrsh} \cite{Stodo}. 
By way of illustration, calculations have been explicitly performed
for Schwarzschild geometry. 

\begin{figure}[t]	
\resizebox{10cm}{!}{\includegraphics{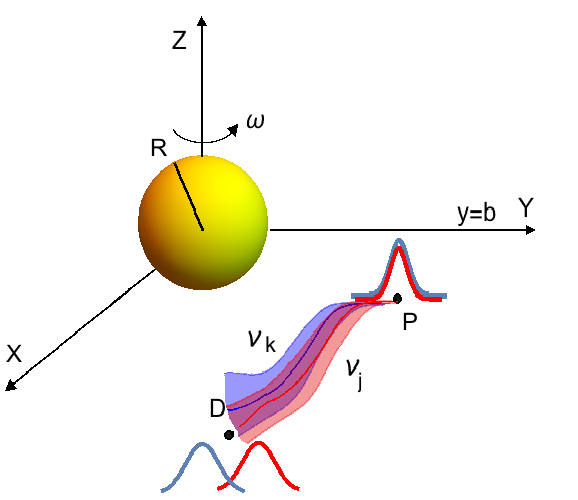}}
\caption{Pictorial representation of a neutrino propagating in the gravitational field of a spinning spherical body from the production point $\mathrm{P}$ to the detection point $\mathrm{D}$, where the WPs can still interfere. We are assuming that each eigenstate WP follows a trajectory close to null-geodesics. The (online) colored widths represent the distribution of the trajectories due to the WP finite extension.}
\label{fig1}
\end{figure}

In curved spacetime a neutrino flavor state produced
at the point $P(t,\vec{x}_P)$ is described by $|\nu_\ell(P)\rangle=\sum_{i=1,2}U^*_{\ell i}|\nu_i(P)\rangle$. During the propagation to the detection point $D(t_{D},\vec{x}_{D})$, 
the $i$-th mass eigenstate evolves according to
\be
\label{CSev}
|\nu_i(P,D)\rangle=e^{-i\Phi_i(P,D)}|\nu_i(P)\rangle\,,
\ee
where the QM phase $\Phi_i(P,D)$ in its covariant form
is given by \cite{Stodo}
\be
\label{Stodolsky}
\Phi_i(P,D)=\int_P^{D} p^{(i)}_{\mu}dx^{\mu}\,.
\ee
Here $p^{(i)}_{\mu}$ ($\mu=0,1,2,3$) is the canonical four-momentum
conjugated to the coordinate $x^\mu$, 
\be
p^{(i)}_{\mu}=m_ig_{\mu\nu}\frac{dx^\nu}{ds}\,,
\ee
satisfying the generalized mass-shell relation
\be
\label{masshell}
p^{(i)}_{\mu}p^{(i)\hspace{0.2mm}\mu}=m_i^2\,,
\ee
where $g_{\mu\nu}$ is the metric tensor and
$ds$ the line element along the trajectory 
described by the $i$-th neutrino state. 
In the approximation of relativistic neutrinos, 
it is reasonable to assume that this trajectory
is close to a null-geodesic.  Clearly, the detection point
$D$ is such that the mass eigenstate WPs can still
interfere in it (see Fig. \ref{fig1}). 

The relation \eqref{CSev} along with the 
definition \eqref{Stodolsky} of the QM phase
generalize Eq. \eqref{coordsp} to an arbitrary curved space.
For $g_{\mu\nu}$ equal to Minkowski tensor, it is easy to check
that the standard evolution in flat spacetime is recovered.

\subsection{The Lense-Thirring metric example}
The Lense-Thirring metric describes the gravitational field around
a spinning spherical source of constant density. Let us denote by $M$
and $R$ the mass and radius of the central source, respectively. 
By assuming the neutrino propagation to be confined to the equatorial 
plane ($z=0$), the line element in the (linearized) weak-field limit
can be written in cartesian coordinates as
\begin{equation}
\label{LTds}
\dr s^2=(1+2\varphi)\dr t^2-\frac{\varphi\hspace{0.2mm}\Omega}{r^2}(x \,\dr y-y\, \dr x) \, \dr t-(1-2\varphi)(\dr x^2+\dr y^2)\, ,
\end{equation}
where 
\be
\Omega\equiv\frac{4R^2\omega}{5}\,,
\ee 
and $\omega$ is the angular velocity of the source, supposed 
to be constant and oriented along the $z$ axis.
The gravitational potential $\varphi\equiv\varphi(r)$ is 
defined by
\be 
\varphi(r)=-\frac{M}{r}\equiv-\frac{M}{\sqrt{x^2+y^2}}\,,
\ee
where, for simplicity, we have denoted the radial distance in the equatorial plane by $r$.

Equation \eqref{LTds} provides the metric typically employed to describe
gravitomagnetic frame-dragging effects \cite{Ruggi,Ciufo}. Moreover, 
in \cite{TEUR} it has been used to compute gravity corrections
to the Mandelstam-Tamm time-energy uncertainty relation for oscillations. 

In the above setting, it is easy to show that the only 
nontrivial components of the four-momentum $p_\mu^{(i)}$
are
\begin{eqnarray}
\label{timeco}
p_t^{(i)}&\hspace{-2mm}=\hspace{-2mm}&m_i\left[(1+2\varphi)\frac{dt}{ds}+\frac{\varphi\hspace{0.2mm}\Omega\hspace{0.2mm}y}{2r^2}\frac{dx}{ds}-\frac{\varphi\hspace{0.2mm}\Omega\hspace{0.2mm}x}{2r^2}\frac{dy}{ds}
\right],\\[2mm]
\label{xco}
p_x^{(i)}&\hspace{-2mm}=\hspace{-2mm}&m_i\left[\frac{\varphi\hspace{0.2mm}\Omega\hspace{0.2mm}y}{2r^2}\frac{dt}{ds}-(1-2\varphi)\frac{dx}{ds}\right],\\[2mm]
\label{yco}
p_y^{(i)}&\hspace{-2mm}=\hspace{-2mm}&m_i\left[-\frac{\varphi\hspace{0.2mm}\Omega\hspace{0.2mm}x}{2r^2}\frac{dt}{ds}-(1-2\varphi)\frac{dy}{ds}\right],
\end{eqnarray}
where we have neglected higher-order terms in the potential $\varphi$.

Following \cite{Ruggi,TEUR}, we can now consider a first approximation
in which the neutrinos propagate along a direction parallel to the  
$x$-axis with impact parameter $y=b>R$ (Fig. \ref{fig1}). This implies that
$dy=0$ in Eqs. \eqref{timeco} and \eqref{yco}. 
Furthermore, since the metric \eqref{LTds} does not depend on $t$, 
the component $p_t^{(i)}\equiv E_i(\vec{p})$ is a constant of motion, 
corresponding to the neutrino energy as measured by an inertial
observer at rest at infinity (asymptotic energy). 
The local energy, which is the
quantity actually measured by an observer
at rest at finite radius in the gravitational field \cite{Fornengo, Cardall, Petruz}, 
is related to $E_i(\vec{p})$ 
trough the transformation law that connects the local Lorentz
frame $\{x^{\hat a}\}$ to the general frame $\{x^{\mu}\}$ 
\be
\label{tetrad0}
x^{\mu}=e^{\mu}_{\hat a} x^{\hat a},
\qquad g_{\mu\nu} e^\mu_{\hat a}e^\nu_{\hat b} = \eta_{\hat a\hat b}\,,
\ee
where $e^\mu_{\hat a}$ are the vierbein fields
and we have used the standard convention
of denoting general coordinates and local Lorentz frame indexes by Greek 
and hatted  Latin  letters, respectively. 
With reference to the metric \eqref{LTds}, the relevant tetrad components
are \cite{TEUR}
\be
\label{tetrad}
e^0_{\hat{0}}=1-\varphi, \quad\,\, e^1_{\hat{0}}=\frac{\varphi\hspace{0.2mm}\Omega\hspace{0.2mm}y}{r^2}, \quad\,\, e^2_{\hat{0}}=-\frac{\varphi\hspace{0.2mm}\Omega\hspace{0.2mm} x}{r^2}, \quad\,\, e^i_{\hat{j}}=(1+\varphi)\delta^i_j\,.
\ee
From Eq. \eqref{tetrad0}, it is a matter of calculation to show that the local and asymptotic energies are related by \cite{TEUR}
\be
\label{locasy}
E_{L}=\lf[1-\varphi\left(1+\frac{\Omega \hspace{0.2mm}y}{r^2}\right)\right]E(\vec{p})\,,
\ee
where the subscript $L$ stands for ``local''
and we have omitted for simplicity the space- and momentum-dependence
of $E_{L}$. 

A comment is now in order: in \cite{Volpe} all the
quantities are expressed in terms of the asymptotic
energy $E_i(\vec{p})$. To make the comparison
with the formulas of \cite{Volpe} easier, in what follows we 
retain the dependence on $E_i(\vec{p})$ and 
implement the substitution \eqref{locasy} only at the end. 

With our assumptions, the mass-shell relation \eqref{masshell} 
takes the form
\be
\label{mass2}
(1+2\varphi)\left(\frac{dt}{ds}\right)^2-(1-2\varphi)\left(\frac{dx}{ds}\right)^2+\frac{\varphi\hspace{0.2mm}\Omega\hspace{0.2mm}b}{r^2}\frac{dt}{ds}\frac{dx}{ds}=1\,, 
\ee
where $r$ must now be intended as $r(x,y=b)$.
From Eq. \eqref{timeco} along with $p_t^{(i)}=E_i(\vec{p})$, we derive
\be
\label{dtds}
\frac{dt}{ds}=\frac{E_i(\vec{p})}{m_i}\left(1-2\varphi\right)-\frac{\varphi\hspace{0.2mm}\Omega\hspace{0,2mm}b}{2r^2}\frac{dx}{ds}\,,
\ee
which can be replaced into \eqref{mass2} to give
\be
\label{dxds}
\frac{dx}{ds}=\pm \frac{E_i(\vec{p})}{m_i}\left[1-\frac{m^2_i}{2E_i^2(\vec{p})}(1+2\varphi)\right].
\ee 
Here we have exploited the condition of relativistic neutrinos at infinity, which ensures
that they are even more relativistic for $r<\infty$, according to Eq. \eqref{locasy}. 
Without loss of generality,  we assume that neutrinos propagate toward increasing values of $x$ as $s$ increases, so that the solution with the positive sign must be considered.

Let us now consider the covariant phase in Eq. \eqref{Stodolsky}.
Combining  Eqs. \eqref{timeco}-\eqref{yco} and \eqref{dtds},\eqref{dxds}, 
one gets for the integral argument
\be
p_\mu^{(i)}dx^\mu=E_i(\vec{p})\left\{dt -\left[(1-2\varphi)-\frac{\varphi\hspace{0.2mm}\Omega\hspace{0.2mm}b}{r^2}-\frac{m_i^2}{2E_i^2(\vec{p})}\right]dx\right\}.
\ee
The covariant phase then reads
\be
\label{phaseres}
\Phi_i(P,D)=E_i(\vec{p}) \left(t_{PD}-{b}_{PD}-c^{(\Omega)}_{PD}\right)+\frac{m_i^2}{2E_i(\vec{p})}x_{PD}\,,
\ee
where, for brevity, we have defined $t_{PD}\equiv t_D-t_P$,  $x_{PD}\equiv x_D-x_P$ and 
\begin{eqnarray}
{b}_{PD}&\hspace{-2mm}\equiv\hspace{-2mm}&x_{PD}+2M\log\left(\frac{x_D+r_D}{x_p+r_P}\right),\\[2mm]
c^{(\Omega)}_{PD}&\hspace{-2mm}\equiv\hspace{-2mm}&\frac{\Omega M}{b}\left(\frac{x_D}{r_D}-\frac{x_P}{r_P}\right), 
\label{comega}
\end{eqnarray}
with 
\be
\label{rdrp}
r_D\equiv r(x=x_D,b)=\sqrt{x_D^2+b^2},\quad\,\,  \,
r_P\equiv r(x=x_P,b)=\sqrt{x_P^2+b^2}.
\ee 
Clearly, for a non-rotating source,  $c^{(\Omega)}_{PD}=0$.
In this case, Eq. \eqref{phaseres} reproduces the linearized 
result of \cite{Volpe} for the Schwarzschild metric, 
up to a global sign due to the different signature 
convention adopted for the metric. 

From Eq. \eqref{phaseres}, the phase difference $\Phi_{jk}=\Phi_j-\Phi_k$ 
of the mass eigenstate WPs reads
\be
\Phi_{jk}(P,D)=\left[E_j(\vec{p})-E_k(\vec{q})\right] \left(t_{PD}-{b}_{PD}-c^{(\Omega)}_{PD}\right)+\left(\frac{m_j^2}{2E_j(\vec{p})}-\frac{m_k^2}{2E_k(\vec{q})}\right)x_{PD}\,.
\ee
By use of the first-order expansion \eqref{Eexpand}, this becomes
\begin{eqnarray}
\label{newphdif}
\Phi_{jk}(P,D)&\hspace{-3mm}=\hspace{-3mm}&E_{jk}\left(t_{PD}-{b}_{PD}-c^{(\Omega)}_{PD}\right)+\left(\frac{m_j^2}{2E_j}-\frac{m_k^2}{2E_k}\right)x_{PD}\\[2mm]
&&+\,\,\vec{v}_j\cdot(\vec{p}-\vec{p}_j)(t_{PD}-{\lambda}_j)
- \vec{v}_k\cdot(\vec{q}-\vec{q}_k)(t_{PD}-{\lambda}_k)\,,
\nonumber
\end{eqnarray}
where we have used Eq. \eqref{diffen}
and introduced the shorthand notation
\be
{\lambda}_i(P,D)\equiv b_{PD}+c^{(\Omega)}_{PD}+\frac{m_i^2}{2E_i^2}x_{PD}\,.
\ee

Let us now evaluate the one-body density matrix
describing the neutrino mass eigenstates as (non-covariant)
Gaussian WPs.  This is given by the generalization
of Eq. \eqref{integ} with the QM phase being given by Eq. \eqref{newphdif}, i.e.
\begin{eqnarray}
\widetilde\rho^{(\ell)}_{jk}(P,D)&\hspace{-1mm}=\hspace{-1mm}&N^{(\ell)}_{jk}\int_{\vec{p},\vec{q}}
e^{-i\Phi_{jk}(P,D)}\,
e^{-\left[\frac{(\vec{p}-\vec{p}_j)^2}{4\sigma_p^2}\,+\,\frac{(\vec{q}-\vec{q}_k)^2}{4\sigma_p^2}\right]}\,, \\[2mm]
\nonumber
&\hspace{-2mm}=\hspace{-2mm}&\frac{N^{(\ell)}_{jk}}{(2\sqrt{\pi}\sigma_x)^6}
\,e^{-iE_{jk}(t_{PD}-b_{PD}-c^{(\Omega)}_{PD})}\,
e^{i\left(\frac{m^2_k}{2E_k}-\frac{m_j^2}{2E_j}\right)x_{PD}}\,
e^{-\sigma_p^2\left[v_k^2\left(t_{PD}-{\lambda}_k\right)^2+v_j^2\left(t_{PD}-{\lambda}_j\right)^2\right]},
\end{eqnarray}
where we have used the tilde to distinguish 
the density matrix in curved spacetime from the
corresponding flat expression.
The normalization $N^{(\ell)}_{jk}$
is defined in Eq. \eqref{norm}. 

As in Sec. \ref{Sec1}, the averaged density
matrix is obtained by integrating $\widetilde\rho^{(\ell)}_{jk}$ over the
coordinate time. Also in this case, the resulting
expression can be factorized into the product of three terms
as follows
\be
\widetilde\rho^{(\ell)}_{jk}(x_P,x_D)= A^{(\ell)}_{jk}\,\widetilde{\rho}_{jk}^{\mathrm{\,\,osc}}(x_P,x_D)\,\widetilde\rho_{jk}^{\mathrm{\,\,damp}}(x_P,x_D)\,,
\ee
to compare with the corresponding
flat result \eqref{factorflat}. The amplitude
$A^{(\ell)}_{jk}$ is independent of the travelled 
distance and exhibits the same expression as
in Minkowski spacetime (see Eq. \eqref{AMP}). 
The second term, which is responsible for flavor
oscillations, is given by
\be
\widetilde\rho_{jk}^{\mathrm{\,\,osc}}(x_P,x_D)=
e^{i\left(\frac{m_k^2}{2E_k}-\frac{m_j^2}{2E_j}\right)x_{PD}}\,
e^{-i\frac{E_{jk}}{v^2}\left(v_j^2\frac{m_j^2}{2E^2_j}+v_k^2\frac{m_k^2}{2E^2_k}\right)x_{PD}}\,.
\ee
Finally, the damping term reads
\be
\widetilde\rho_{jk}^{\mathrm{\,\,damp}}(x_P,x_D)=
e^{-\frac{(v_j v_k x_{PD})^2}{4v^2\sigma_x^2}\left(\frac{m_k^2}{2E^2_k}-\frac{m_j^2}{2E^2_j}\right)^2}\,, 
\ee
which for relativistic neutrinos becomes 
\be
\label{rhoapp}
\widetilde\rho_{jk}^{\mathrm{\,\,damp}}(x_P,x_D)\simeq e^{-\frac{(m_k^2-m_j^2)^2\hspace{0.2mm}}{32 E^4\sigma_x^2}x_{PD}^2}.
\ee
Here $E$ is the average energy between the
mass eigenstates, as defined below Eq. \eqref{cohle}.

We notice that Eq. \eqref{rhoapp} is formally the same as the
damping term found in \cite{Volpe} in Schwarzschild metric.
As argued in \cite{Petruz}, to make the dependence
on $M$ and $\Omega$ explicit, $\widetilde\rho_{jk}^{\mathrm{\,\,damp}}$
should be recast in terms of the local (rather than asymptotic)
energy \eqref{locasy}. By implementing the transformation \eqref{locasy}
into Eq. \eqref{rhoapp}, we are led to
\be
\label{rhodampCS}
\widetilde\rho_{jk}^{\mathrm{\,\,damp}}(x_P,x_D)\simeq e^{-\frac{(m_k^2-m_j^2)^2\hspace{0.2mm}x_{PD}^2}{32\,E_L^4\sigma_x^2}\,
\left[1+\frac{4M}{r_D}\left(1+\frac{\Omega\hspace{0.2mm}b}{r_D^2}\right)
\right]},
\ee
where the local energy must be considered as being
evaluated at the detection point $D$.

In analogy with the flat-spacetime case, 
one can now define the proper coherence length
as the distance at which the density matrix is
suppressed by the factor $e^{-1}$. We then obtain
\be
\label{final}
\widetilde L^{PD}_\mathrm{coh}\simeq\frac{4\sqrt{2}\hspace{0.2mm}E_L^2\hspace{0.2mm}\sigma_x}{|\Delta m_{jk}^2|}
\left(1-2\frac{M}{r_D}
-2\frac{M\hspace{0.2mm}\Omega\hspace{0.2mm}b}{r_D^3}\right).
\ee
We remark that in the $M\rightarrow0$ limit, 
Eq. \eqref{cohleDM}
is straightforwardly recovered, since $E_L\rightarrow E$. 
This is indeed the expected outcome
in the absence of gravitational field. 
Therefore, respect to the flat case, 
the coherence length in the linearized Lense-Thirring metric 
turns out to be decreased by acquiring some nontrivial corrections.
Specifically, the second term in the brackets 
is a constant factor that only depends
on the source mass and the detection point. 
The presence of this correction could be somehow expected, 
since the spacetime metric is no longer
translation-invariant. The third term is the genuinely 
Lense-Thirring imprint, as it carries the information
about the rotational velocity $\Omega$ of the central object
and the impact parameter $b$ of neutrino's trajectory.

\begin{figure}[t]	
\resizebox{13.5cm}{!}{\includegraphics{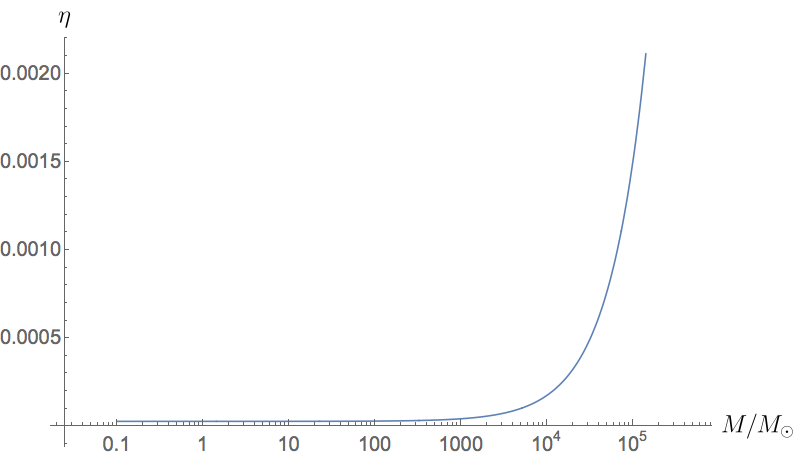}}
\caption{Plot of the relative correction $\eta$ versus the source mass
$M$ (in solar mass units). The logarithmic scale is used on 
the horizontal axis.}
\label{fig2}
\end{figure}

It would be interesting to estimate the gravity corrections in Eq. \eqref{final}. 
For this purpose,  we suppose to consider neutrinos of
typical energy $E_L\simeq 0.5\, \mathrm{MeV}$ and mass difference
$\Delta m_{jk}^2\simeq 10^{-5}\, \mathrm{eV}^2$.
For the WP width, we take $\sigma_{x}\simeq 10^{-12}\,\mathrm{cm}$ 
as in \cite{Matter2,Matter3,Volpe}. By assuming the mass of the Sun $M_{\odot}\simeq 10^{30}\,\mathrm{Kg}$ and its rotational frequency $\omega^{-1}\simeq 27\,\mathrm{d}$ as sampling values
for the gravity source and setting $b\gtrsim R_\odot\simeq 10^{8}\,\mathrm{m}$, $r_D\simeq 10^{11}\,\mathrm{m}$, we obtain the following estimate 
for the relative difference between the coherence length \eqref{final} and the
corresponding flat-spacetime expression 
\be
\label{relatdif}
\eta\equiv \frac{|\widetilde L^{PD}_{\mathrm{coh}}-\widetilde L^{PD}_{\mathrm{coh}}(M=0)|}{\widetilde L^{PD}_{\mathrm{coh}}(M=0)}\simeq 10^{-8}\,,
\ee
thus showing the negligible impact of  Sun's gravity
on decoherence effects, as predictable for very weak fields. 
However, for neutrinos in astrophysical environments, 
such as supernova neutrinos or neutrinos propagating
in the field of massive compact objects, 
these effects are expected to become
significant. For instance, in Fig. \ref{fig2} the 
relative difference \eqref{relatdif} is plotted as a function of $M$
and for distances $r_D$ such that the weak-field approximation
is still valid. 
This shows that the 
contribution of gravity corrections becomes increasingly relevant
as one approaches strong-gravity regimes.
Of course, a more rigorous 
analysis of decoherence effects in this scenario
would require us to go beyond the linearized
approximation. This study is under active
investigation and will be presented elsewhere \cite{In prepa}.

Finally, in order to compare Eq. \eqref{rhodampCS} with the result of \cite{Petruz}, 
we need to express the coordinate distance $x_{PD}$ 
in terms of the proper distance $L_{PD}$ between the production and detection points. Under our assumptions, it is possible to show that these two quantities are related by \cite{TEUR}
\be
L_{PD}=\int_P^D \sqrt{-g_{xx}}dx= 
x_{PD}+M\log\left(\frac{x_D+r_D}{x_P+r_P}\right),
\ee
where $r_D$ and $r_P$ are defined in Eq. \eqref{rdrp}. 
By plugging into Eq. \eqref{rhodampCS}, 
the averaged density matrix takes the form
\be
\widetilde\rho_{jk}^{\mathrm{\,\,damp}}(x_P,x_D)\simeq e^{-\frac{(m_k^2-m_j^2)^2\hspace{0.2mm}L_{PD}^2}{32\,E_L^4\sigma_x^2}\,
\left[1+\frac{4M}{r_D}\left(1+\frac{\Omega\hspace{0.2mm}b}{r_D^2}\right)
-\frac{2M}{L_{PD}}\log\left(\frac{x_D+r_D}{x_P+r_P}\right)
\right]},
\ee
which gives rise to the modified coherence length 
\be
\label{colebis}
\widetilde L^{PD}_\mathrm{coh}\simeq\frac{4\sqrt{2}\hspace{0.2mm}E_L^2\hspace{0.2mm}\sigma_x}{|\Delta m_{jk}^2|}
\left[1-2\frac{M}{r_D}
-2\frac{M\hspace{0.2mm}\Omega\hspace{0.2mm}b}{r_D^3}\right]
+M\log\left(\frac{x_D+r_D}{x_P+r_P}\right)\,.
\ee

For vanishing spinning velocity $\Omega=0$, it is easy to check that
the above formula reproduces the coherence length obtained
in \cite{Petruz} for the case of a non-rotating Schwarzschild source\footnote{Strictly speaking, the 
second term in the square brackets
is found with opposite sign
with respect to \cite{Petruz}, due to a possible mistake therein.}.
By comparison with Eq. \eqref{final}, 
it follows that $\widetilde L^{PD}_\mathrm{coh}$
acquires a further correction with the
same logarithmic behavior  
as in Schwarzschild spacetime \cite{Petruz}. 
We plan to further investigate the physical meaning of Eq.~\eqref{colebis} 
in order to understand how to express the quantities in terms of proper distance in a more consistent way.


\section{Discussion and Conclusions}
The influence of
gravity on neutrino decoherence has been investigated within
the framework of the density matrix with Gaussian WPs.
As a specific background, we have considered
the gravitational field around a spinning spherical body described 
by Lense-Thirring metric. By working in the weak-field,
we have derived the effective coherence length for relativistic neutrinos, 
showing that it is nontrivially
modified with respect to the flat case.
A rough estimation of gravity corrections has highlighted 
that they are below the sensitivity of current experiments
for neutrinos propagating in the gravitational field of the Sun.
However, significant deviations from the standard
result are expected in strong-field regimes, e.g.
in the case of neutron stars formed from a core-collapse
supernova \cite{Matter2,Volpe} or in the presence of 
supermassive black holes. This is in line with \cite{Volpe}. 

We remark that our study relies on
some preliminary assumptions. For instance, 
we have neglected matter and neutrino self-interactions
outside the central compact object. As claimed
in \cite{Volpe}, these effects can be embedded 
by using a similar procedure involving the use of the matter
eigenstate basis instead of the mass one. 
This investigation turns out to be necessary
to explore whether WP decoherence suppresses
flavor oscillations and its impact on the supernova
dynamics and $r$-process nucleosynthesis. 
Furthermore, we have considered rectilinear propagation
of mass eigenstate WPs. In a more general
treatment, bending effects should also be
contemplated. To give a rough estimation
of the correction we have neglected, let us think
of neutrinos as nearly massless particles. 
In this case, we can refer to \cite{Ruggi,Cohen}, 
where it has been shown that the magnitude
of the bending angle $\delta\phi$ for a light ray in 
Lense-Thirring metric and in the equatorial
plane turns out to be
\be
\delta\phi\simeq \frac{M}{b} \left(1-\frac{\vec{J}\cdot\vec{n}}{M\hspace{0.2mm} b}\right),
\ee
where $\vec{J}$ is the angular momentum 
of the source and $\vec{n}$ is a unit vector
in the direction of the angular momentum of
light about the center of the source-body. The term
outside the parenthesis provides the pure Schwarzschild
correction. For instance, in the case of the Sun, 
the relative correction to the pure mass term
is of order of $10^{-6}$, which in principle justifies
our assumption.  

Apart from their intrinsic relevance, let us
emphasize that decoherence effects on neutrino oscillations
are also studied to constrain quantum gravity models \cite{Mavromatos, Nicolini, Test}. In particular, in \cite{Mavromatos} quantum gravitational
analogues of the MSW effect and of foam models endowed
with stochastic fluctuations of the background 
are presented as possible alternative sources of
decoherence in neutrino oscillations. A similar analysis
has been recently developed in \cite{Stutt}, where
the influence of quantum gravity on neutrino 
propagation and decoherence has been investigated
with a focus on the case of neutrino
interactions with virtual black holes produced by spacetime fluctuations.  
The study of the above aspects is quite demanding
and will be object of future works \cite{In prepa}. 

\label{Sec3}







\end{document}